**EUGENE GARFIELD, FRANCIS NARIN, AND PAGERANK: THE THEORETICAL BASES OF THE GOOGLE SEARCH ENGINE**

by

Stephen J. Bensman
LSU Libraries
Louisiana State University
Baton Rouge, LA 70803   USA
E-mail address: notsjb@lsu.edu

**Abstract**

This paper presents a test of the validity of using Google Scholar (GS) to evaluate the publications of researchers. It does this by first comparing the theoretical premises on which the GS search engine PageRank algorithm operates to those on which Garfield based his theory of citation indexing. It finds that the basic premise is the same, i.e., that subject sets of relevant documents are defined semantically better by linkages than by words. Google incorporated this premise into PageRank, amending it with the addition of the citation influence method developed by Francis Narin and the staff of Computer Horizons, Inc. (CHI). This method weighted more heavily citations from documents which themselves were more heavily cited. Garfield himself essentially had also incorporated this method into his theory of citation indexing by restricting as far as possible the coverage of the *Science Citation Index* (SCI) to a small multidisciplinary core of journals most heavily cited. From this perspective, PageRank can be considered a further implementation of Garfield's theory of citation index at a higher technical level. Stealing a page from Garfield's book, the paper presents a test of the validity of GS by tracing its citations to the h-index works of 5 Nobel laureates in chemistry—the discipline in which Garfield began his pioneering research—with Anne-Wil Harzing's revolutionary Publish-or-Perish (PoP) software that has established bibliographic and statistical control over the GS database. Most of these works were journal articles, and the rankings of the journals in which they appeared by both total cites (TC) and impact factor (IF) at the time of their publication were analyzed. The results conformed to the findings of Garfield through citation analysis, confirming his law of concentration and his view of the importance of review articles. As a byproduct of this finding, it is shown that Narin had totally misunderstood and mishandled citations from review journals. The evidence of this paper is conclusive: Garfield's theory of citation indexing and PageRank validate each other, and Eugene Garfield is the grandfather of the Web search engine.

**Introduction**

This paper is an investigation into the historical origins of the Google search engine, which is based on an algorithm called PageRank. It is part of a research project aimed at determining whether Google Scholar (GS) can be utilized to evaluate the publishing performance of researchers. The paper shows that PageRank is further development at a higher technological level of the theory of citation indexing developed by Eugene Garfield, founder of the Institute for Scientific Information (ISI), which launched the *Science Citation Index* (*SCI*), the *Social Sciences Citation Index* (*SSCI*), and the *Arts & Humanities Citation Index* (*A&HCI*). ISI is now a unit of Thomson Reuters. The paper begins by describing the historical breakthrough made by Eugene Garfield in uncovering the structure of scientific literature in his development of citation indexing. Garfield did a number of analyses of the citation patterns to the works of Nobelists as a way of validating citations as a quantitative measure of the quality and importance of scientific work. Concomitantly he developed two citation measures for journals—total citations (TC) and the impact factor (IF)—that allowed the determination of which journals are most likely to publish important scientific work. Of these measures he found that TC best identified the important research journals, and this measure lay at the basis of his Law of Concentration, which posited that the scientific journal system is dominated by a small multidisciplinary core of as few as 500 but no more than 1,000 highly cited journals. On the other hand, Garfield found that IF was uniquely able to pinpoint the important review journals, where the paradigms of science are defined. The paper then shows how the two founders of Google, Lawrence Page and Sergey Brin, incorporated the basic principles of citation indexing as developed by Garfield in PageRank, which became the foundation of Google's search ranking algorithm. This incorporation was not direct but as these principles had been modified by the "influence" method devised by Francis Narin and the staff of Computer Horizons, Inc., whereby citations from heavily cited journals were weighted more than citations from less cited journals. The "influence" method was utilized in the first assessment of U.S. research-doctorate programs that employed not just peer ratings but also publication measures. In its explanation of both citation indexing and the Google search engine the paper emphasizes that the initial motivation of their creators in linking documents was not the evaluation of researchers but the creation of coherent or relevant document sets in response to search queries. The paper concludes by analyzing the frequency distributions of the hyperlinks to the works of 5 Nobel laureates in chemistry, which were downloaded from Google Scholar with Anne-Wil Harzing's revolutionary Publish-or-Perish (PoP) software that has established statistical control over this database. The results were fully consistent with Garfield's findings on the structure of the scientific journal system through citation analysis, thereby validating the use of Google Scholar to evaluate the publishing performance of researchers. These results are not surprising as both citation indexing and PageRank are based upon the same theoretical premises

**The Historical Breakthrough: Eugene Garfield and Citation Indexing**

Eugene Garfield is the creator of citation indexing. In his landmark book on the subject Garfield (1983) gave the following conceptual definition of citation indexing:

> The concept of citation indexing is simple…. Citations are the formal, explicit
>
> linkages between papers that have particular points in common. A citation index
>
> is built around these linkages. It lists publications that have been cited and identifies
>
> the sources of the citations. Anyone conducting a literature search can find from
>
> one to dozens of additional papers on a subject just by knowing one that has been cited.
>
> And every paper that is found provides a list of new citations with which to continue
>
> the search. (p. 1)

In an article entitled "Citation Indexes for Science" published in the journal *Science* Garfield (1955) set forth the basic reasons for developing a citation index. Later in life Garfield (1987a) deemed this article "my most important paper" (p. 16). In his *Science* article Garfield (1955) stated that a primary advantage of a citation index over conventional alphabetical and subject indexes was that its different construction allowed it to bring together material that would never be collated by the usual subject indexing. Garfield here described a citation index as "an association-of-ideas index" (p. 108) that allowed the reader as much leeway as he needed. In his opinion, conventional indexes were inadequate, because scientists were often concerned with a particular idea rather than a complete concept, and the basic problem was to build subject indexes that can anticipate the infinite number of possible approaches that scientists may require in order to bridge the gap between the subject approach of those who create the documents and the subject approach of those who seek the information. Garfield stated that the utility of a citation index had to be considered from the viewpoint of the transmission of ideas. Thus, Garfield justified citation indexing as better able to deliver a set of relevant documents in response to a scientist's search query.

To implement citation indexing, Garfield established a company, which he called the Institute for Scientific Information (ISI). ISI began regular publication of the *Science Citation Index* (*SCI*) in 1964. It introduced the *Social Sciences Citation Index* (*SSCI*) in 1973 and the *Arts & Humanities Citation Index* (*A&HCI*) in 1978. Garfield had an intense interest in the history and sociology of science, and these indexes provided invaluable data for pursuing this interest. He entered a collaborative arrangement with leading scholars, who also resorted to this data. One was Derek J. de Solla Price, who was the Avalon Professor of the History of Science at Yale University and is regarded as the father of scientometrics. Another was Robert K. Merton, who is generally considered the founder of the sociology of science. Merton taught for most of his career at Columbia University, where Garfield obtained his degrees in chemistry and library science, In the mid-1960s Merton's interest in science as an exemplar of sociological theory caused him to establish a Columbia research program in the sociology of science, to which he recruited as its first students Jonathan Cole, Stephen Cole, and Harriet Zuckerman—all of whom became important sociologists of science in their own right. Zuckerman (1977) wrote a classic study of Nobel laureates in the United States, in which she utilized *SCI* citation data, and it was on the basis of her research that Merton (1968) developed his concept of the "the Matthew Effect" as descriptive of the psychosocial processes affecting the allocation of awards to scientists for their contributions. He derived this concept from the Gospel according to Matthew (13:12 and 25:29), which states in the King James translation he preferred: "For unto every one that hath shall be given, and he shall have abundance; but from him that hath not shall be taken away even that which he hath." Merton considered the Matthew Effect as causing a complex pattern of the misallocation of credit for scientific work, and he summed up this misallocation thus: "…the Matthew effect consists in the accruing of greater increments of recognition for particular scientific contributions to scientists of considerable repute and the withholding of such recognition from scientists who have not yet made their mark" (p. 58).

Analysis of the publication and utilization patterns of the works of Nobelists is a viable way to test the validity of a scientometric measure. Established by the will of the Swedish chemist, Alfred Nobel, the inventor of dynamite, this prestigious prize has been awarded every year since 1901 for achievements in physics, chemistry, physiology or medicine, literature, as well as for peace. It is an international award administered by the Nobel Foundation in Stockholm. In 1968 the Sveriges Riksbank, Sweden's central bank, established The Sveriges Riksbank Prize in Economic Sciences in Memory of Alfred Nobel. Thus, one is able to analyze the publication and utilization patterns in three different sciences, one social science, and literature. Defining its importance, Zuckerman (1977) wrote that "…the prize has become the supreme symbol of excellence in science, and to be a Nobel laureate is, for better or worse, to be firmly placed in the scientific elite" (p. xiii). Garfield began utilizing Nobelists to promote his product almost immediately after the *SCI* began regular publication. At a 1965 conference sponsored by the Office of Naval Research on research program effectiveness, Sher and Garfield (1966) tallied the citations in the 1961 *SCI* to the persons awarded the Nobel Prize in physics, chemistry, and medicine in 1962 and 1963. They then calculated the number of citations per prize winner, finding that as a group there were 30 times as many citations per average Nobelist as there were per average cited author in the 1961 *SCI* database. Their conclusion was that "there are a number of different methods and occasions for the use of citation data in evaluating and improving the effectiveness of research" (p. 142).

Over the years numerous studies relating to Nobelists were conducted at ISI. Garfield and Welljams-Dorof (1992) reviewed six of these studies, which they deemed major, in response to the invitation of a journal editor to contribute to a special issue devoted to the Nobel Prize. Welljams-Dorof was Garfield's scientific assistant at ISI. Garfield and Welljams-Dorof (1992) defined the purpose of their study as "… to review six major ISI rankings of high impact authors to determine how many already were Nobel laureates at the time each ranking was published and how many authors later went on to win the prize" (p. 118). Overall they found that in the highest percentile of cited researchers, e.g. the top 0.1%, a significant percentage had either won the Nobel Prize or went to win the prize in later years, leading them to the conclusion that author citation rankings are an effective method for identifying both past and present Nobelists as well as future laureates.

Of the six ISI rankings reviewed, two are of especial interest here: 1) the 250 most-cited primary authors, 1961-1975 (Garfield, 1977); and 2) the 300 most-cited authors, 1961-1976, including co-authors (Garfield, 1978b). The latter reported the number of citations as primary author and the number of citations as co-author, combining the two for a total. These rankings are of interest here, because Google Scholar retrieves documents attributable to a given researcher no matter what the position of the researcher in the authorship structure—be it primary, secondary, or even editor—and the position of the Nobelists in the authorship structure was found to be highly variable. The two lists are not directly comparable. For example, the primary-author list had no time limitation on what was being cited and covered not only journal articles but books, whereas the list including co-authors was restricted to journal articles published since 1961. This can be considered part of the reason that only 77 researchers were common to both lists. There is a discrepancy between the number of Nobelists stated in the original studies and what was reported by Garfield and Welljams-Dorof, so both sets of numbers will be given here. According to the original study (Garfield, 1977, Part II, p. 337), of the 250 most-cited primary authors, 42 or 17% were Nobel Prize winners: 15 in physiology or medicine; 14 in physics; and 13 in chemistry. Garfield and Welljams-Dorof (1992, p. 119) report that there were 38 Nobelists plus 13 researchers, who were later to win the prize, for a total of 51 Nobelists or 20%. As for the 300 most-cited authors including co-authors, the original study (Garfield, 1978b, Part 3, p. 588) reported that 26 or 9% had won the prize: 18 in physiology or medicine, 6 in chemistry, and 2 in physics. However, the study stated that 11 of these Nobelists did not make primary author list. According to Garfield and Welljams-Dorof (1992, p. 119), of the 300 most-cited authors including co-authors, 22 had won the prize, 15 were to win the prize, for a total of 37 or 12%. Taken together, the figures indicate that Nobelists publish a large part of their work as secondary authors. This is particularly evident in the figures of Garfield and Welljams-Dorof, where 13 of the 51 Nobelists or 24% on the primary author list were yet to win the prize, and the corresponding numbers on the list with co-authors were 15 of the 37 Nobelists or 41%.

Besides the sciences, Garfield and Welljams-Dorof (1992, pp. 129-130) also took up the question of how well citations predicted Nobelists in economics and literature. In respect the first, they referred to a study done at ISI two years earlier (Garfield, 1990). This study ranked the 50 most-cited economists in the *SSCI* for the years 1966-1986. The citations were attributed only to primary authors and included citations not only to journal articles but also books. Of the 50 most-cited economists 15 or 30% were found to be Nobelists, accounting for an incredible 62.5% of the 24 economics awards granted through 1986, Garfield and Welljams-Dorof report that two laureates-to-be were also listed. It was pointed out to Garfield (1990, p. 5) that the exclusion of secondary authors from the count had probably more than halved the citations to Joseph E. Stiglitz, who appeared on the list but did not win the prize until 2001. As for literature, Garfield and Welljams-Dorof (pp. 129-130) stated that citations also function well as predictors here despite all the controversies about personal, geographic, and philosophical biases. For proof of this, they referred to a study done shortly after the *A&HCI* began publication of the 100 authors eligible for the prize and most cited in this index during 1977-78 (Garfield, 1980b). Of these authors, 23 were indicated as Nobelists, and 2 others were laureates-to-be.

Garfield made his historical breakthrough on the probabilistic structure of journal literature in an ISI study that led to the creation of the *Science Citation Index Journal Citation Reports* (*SCI JCR*). The *JCR* is an annual publication containing citation as well as other quantitative data on the journals covered by it. There are two *JCR*s—one for the *Science Citation Index*, which began publication in 1975, and one for the *Social Sciences Citation Index*, which appeared in 1977. The study was conducted in 1971, and it was based upon all the references published during the last quarter of 1969 in the 2,200 journals then covered by the *Science Citation Index* (*SCI)*. Garfield (1972a) summarized the results of this project in a landmark article published in the journal *Science*, and he periodically reported on the progress of this study in *Current Contents*, ISI's alerting service for journal articles. Two citation measures were utilized to analyze the structure of the scientific journal system—total cites (TC) and the impact factor (IF). It was by analyzing the ranking of journals by TC that Garfield solved the problem posed by Bradford's Law of Scattering for the indexing of scientific journal literature. S. C. Bradford was head of the Science Museum Library in London, one of whose goals was improving the indexing and abstracting of scientific journals. In pursuing this goal he had an analysis conducted of the distribution of articles on a given subject over journals. On the basis of this analysis Bradford (1934) derived his Law of Scattering, to which he gave the following verbal formulation:

> ...if scientific journals are arranged in order of decreasing productivity of articles
> on a given subject, they may be divided into a nucleus of periodicals more particularly devoted to the subject and several groups or zones containing the same number of articles as the nucleus, when the number of periodicals in the nucleus and succeeding zones will be as $1 : n : n^2$.... (p. 86).

Bradford's Law leads to a highly skewed distribution of articles on a given subject across journals with a small proportion of the journals accounting for most of the articles. Thus, Bradford's figures for applied geophysics are summarized by Bensman (2001, p. 239) in the following manner: nucleus -9 journals (2.8%) accounting for 249 (32.2%) articles; first zone – 59 (18.1%) journals accounting for 499 (37.5%) articles; and second zone – 258 (79.1%) journals accounting for 404 (30.3%) articles. Analyzing the problem posed by Bradford's Law for the indexing of science, Garfield (1971) noted in *Current Contents* that this law dictated that "no matter what the specialty, a relatively small core of journals will account for as much as 90% of the significant literature, while attempts to gather 100% of it will add journals to the core at an exponential rate." He then stated, "Any abstracting or indexing service that ignores Bradford's law in attempting to realize the myth of complete coverage does so at its great financial peril." (p. 5).

In analyzing scientific journals in terms of TC Garfield (1972a) found the same type of distributions as Bradford had found in the distribution of articles on a given subject across journals. He described these findings in the following manner:

[The plot of the distribution of citations among cited journals] shows that only

25 journals (little more than 1 percent of *SCI* coverage) are cited in 24 percent of

all references; that only 152 journals…are cited in 50 percent of all references; that

only 767 journals are cited in 75 percent of all references; and that only 2000 or

so journals are cited in 85 percent of all references. In addition, the data…show

that only 540 journals are cited 1000 or more times a year, and that only 968 journals

are cited even 400 times a year. (p. 474)

He also demonstrated that this same pattern held true for the distributions of journals by number of articles and number of references to other journals. According to his figures, of the 2200 journals covered by the *SCI* in 1969, about 500 published approximately 70 percent of all the articles, and another small group of 250 journals provided almost half of the 3.85 million references processed for the *SCI* in 1969.

These figures caused Garfield to conclude that "many journals now being published seem to play only a marginal role, if any, in the effective transfer of scientific information" (p. 475).

The structure of *SCI* data caused Garfield (1972a) to observe, "The predominance of cores of journals is ubiquitous" (p. 475). This observation marked his greatest theoretical breakthrough and solution to the problem posed by Bradford's Law of Scattering for abstracting and indexing services. Garfield (1971) announced this breakthrough in *Current Contents* thus:

> …At ISI, we are completing a study which has resulted in a generalization of Bradford's law which, in a sense, "unifies" the demonstration of its validity in studies of individual fields. Allow me the eponymic shorthand of calling this unified theory or generalization "Garfield's law of concentration." The name is intended to suggest that, in opposition to scattering, a basic concentration of journals is the common core or nucleus of all fields. (p. 5)

He described his law as postulating for science as a whole what Bradford's law had postulated for a single discipline, declaring that his law held true no matter whether journals were considered as a source of citing articles or as a collection of cited articles. Garfield (1972a) summarized his law in his *Science* article thus:

> The data reported here demonstrate the predominance of a small group of journals in the citation network. Indeed, the evidence seems so conclusive that I can with confidence generalize Bradford's bibliographical law concerning the concentration and dispersion of the literature of individual disciplines and specialties. Going beyond Bradford's studies, I can say that a combination of the literature of individual disciplines and specialties produces a multidisciplinary core for all of science comprising no more than 1000 journals. The essential multidisciplinary core could, indeed, be made up of as few as 500 journals…. (p. 476)

In his magisterial monograph on citation indexing Garfield (1979, pp. 21-23 and 160) used as a physical analogy for Bradford's law a comet, whose nucleus represents the core journals of a literature and whose tail of debris and gas molecules widening in proportion to the distance from the nucleus depicts the additional journals that sometimes publish material relevant to the subject. According to him, his Law of Concentration postulates that the tail of the literature of one discipline largely consists of the cores of the literatures of other disciplines.

Garfield reinforced his Law of Concentration with other studies. In one such study Garfield (1973) analyzed the frequency with which a given journal publishes articles that become citation superstars. In doing so, he used Oliver Lowry's paper on protein determination, published in 1951 in the *Journal of Biological Chemistry*, as an example of a citation superstar, noting that in 1969 only about 150 journals were cited as frequently as this single paper. Garfield then reported that he had recently examined a list of the 1000 papers most frequently cited during the past decade, finding that only about 200 journals had accounted for these 1000 articles, of which half had been published in only 15 journals. According to Garfield, whereas a citation blockbuster like Lowry's paper was atypical of most scientific papers, it was not so atypical of other papers published in the *Journal of Biological Chemistry*. Summing up his main point on this matter, Garfield wrote that "it is remarkable that of the 1000 or more most heavily cited articles in the literature, not one appeared in an 'obscure' journal" (p. 6). Of the 250 most-cited primary authors in 1961-1975, Garfield (1977, Part 3, p. 6) found that just 17 journals accounted for over half of their most cited articles and that less than 100 journals accounted for all of them. Garfield (1978b, Part 3A, p.15) discovered that the same phenomenon held true for the 300 most-cited authors including co-authors in 1961-1976. Their 300 most-cited articles appeared in only 86 journals, of which 5 journals published more than a third, and ten journals accounted for about half.

From the probabilistic viewpoint, two studies, done by Garfield in the 1990s, are of particular interest for revealing the extraordinary stability of the scientific journal system. This indicates there may be constant probabilities operating over time. In the first Garfield (1991) compared the top 50 journals most highly cited in 1969 to the top 50 journals listed as most highly cited in the 1989 *SCI JCR*, finding very little change. In the second paper, which was specifically devoted to the "concentration" effect, Garfield (1996) compared data derived from the 1987 and 1994 *SCI JCR*s and found that the distributional characteristics of journals by citations received and articles published remained virtually the same with a small proportion of the journals accounting for the bulk of both. To analyze further this phenomenon, he constructed two tables which compared the top 50 journals most highly cited and most productive in articles in 1989 and 1994. In terms of TC Garfield found that the rankings in both years were virtually the same, with 48 journals being on both lists. He found a somewhat lower stability in terms of articles published with 39 journals being on the 1989 and 1994 lists.

The impact factor (IF) was created by Garfield to correct what he considered the two major faults of TC—the effect of physical size and the effect of temporal size. In the first paper, where "impact factor" was formulated in the form ISI was to employ it, Garfield and Sher (1963) described ranking journals by TC as "not much more sophisticated than ranking the importance of a journal by the quantity of articles published," declaring, "The first step in obtaining a more meaningful measure of importance is to divide the number of times a journal is cited by the number of articles that journal has published" (p. 200).. In other words, they posited the arithmetic mean of citations per article as the true measure of journal quality.  The problems and effect of ranking journals by mean citation rate per article were carefully analyzed in the 1971 ISI study of 1969 *SCI* citations that led to the creation of the *JCR*.  In his *Science* article summarizing the results of this study Garfield (1972a) introduced this analysis with the comment, "I have very rarely found among the 1000 most frequently cited journals one that is not also among the 1000 journals that are most productive in terms of articles published."  The conclusion, which he drew from this fact was stated thus:  "In view of the relation between size and citation frequency it would seem desirable to discount the effect of size when using citation data to assess a journal's importance…" (p. 476).  Garfield (1972b) discussed the method, reasons, and results of the impact factor analysis most clearly and succinctly not in the *Science* article but in a *Current Contents* essay.  Here he introduced the problem of temporal size by warning that, although the ratio of citations to sources provides an overall measure of impact, this ratio can be skewed by a few super-cited classics unless limited by chronological criteria (p. 5).  Garfield (1996, p. 13) demonstrated that the effect of a citation classic can be significant by pointing out that, of the 265,000 citations in 1994 to the top-ranked *Journal of Biological Chemistry*, 7,000 or 3 % were accounted for by Lowry's 1951 protein determination paper.  In his *Current Contents* essay on the impact factor Garfield (1972b) described the solution for this fault devised by the ISI project analyzing the 1969 citations thus:

> … we have computed impact factors by dividing the citations to the years 1967 and 1968
>
> by the total number of articles published in 1967 and 1968.  The citations counted appeared
>
> in the 1969 literature. In this way we have chosen a current impact factor which discounts
>
> the effect of most superclassics.  (p. 6)

With this there was established the basic method of the IF to correct for the effects of the physical and temporal size on TC—estimate the arithmetic mean of the citations to the articles of a journal within a two-year time limit, equalizing by both physical size and an arbitrary time limit—and the major reason for limiting by time was to nullify the effect of the citation classics.  However, it should be noted here that the citation classics comprise the historical significance of the journals of the TC citation core of Garfield's Law of Concentration and a major reason for the stability of this core over time.   Garfield (1972b) summarized the results of these corrections by comparing two ranked lists of 50 journals—one by TC and the other by IF—thus:

> Two things are immediately apparent. First, the list of 50 journals with the highest impact, current or otherwise, is quite different from the list of 50 most-cited journals. Only eleven journals appear on both lists. Second, almost half of the high-impact journals can be called review journals. None of them appears on the top 50 most-cited list …. (p. 6)

Thus, it was found at the very beginning that a distinguishing characteristic of the impact factor at least in the sciences was its ability to identify small review journals and bring them to the top of the rankings.

Of the two citation measures—TC and IF—it was IF that came to be regarded as by far the most important. One reason for this is the way the *JCR*s presented the data. Starting with the 1979 *SCI JCR* journals were ranked within subject categories by IF to enable "the *JCR* user to analyze citation data within subject categories" (Garfield, 1980a, p. 21A). This was the only way journals were ranked within subject categories, and, throughout the years of the pre-digital rigid formats of paper and microfiche, this was the only way users could easily judge how journals ranked within subject categories. A reason for this arrangement must be sought in the importance Garfield assigned to review articles. For Garfield, the review article played an important role in the defining of scientific paradigms. Here we are defining "paradigm" in the sense given it by Kuhn (1970) as "scientific achievements … that some particular scientific community acknowledges for a time as supplying the foundation for its further practice" (p. 10). Concerning the review article Garfield (1987b) once stated:

> … It is not an accident that so many of our greatest scientists have used, created, and contributed to the review literature. Like an important opinion rendered by the chief justice of the Supreme Court, reviews can have great value and influence…. (p.5)

Garfield's emphasis on the importance of the review article can be traced back to the influence of two persons, who can be considered his mentors. The first was the radical British scientist, J.D Bernal (1940), whose book, *The Social Function of Science*, was virtually Garfield's intellectual bible in his youth, playing a crucial role in the formation of his thought. In this book Bernal (pp. 297-298) stressed the importance of review literature in scientific communication. Here he recommended that that the responsible body in each scientific discipline periodically review its field and report for each period what it deemed to be the chief discoveries and improvements in its subject.  Bernal also desired that qualified scientists be persuaded to write up their fields at suitable intervals in monographs and textbooks, suggesting as a model for such works the monumental series of German *Handbücher.*  The other person bringing the importance of the review article to Garfield's attention was Chauncey D. Leake, a polymath, who became a president of the American Association for the Advancement of Science.  Early in his intellectual development, Garfield (1970; 1978a) was urged by Leake to study review articles and try to understand why they were so important in science.  Garfield did so and, in essence, created the citation indexing of science by integrating the structure of the review article with that of Shepard's Citations, a citator or index employed in American common law that lists of all the authorities citing a particular case, statute, or other legal authority.  From this perspective it must be posited that a major reason for the *JCRs* ranking journals within subject categories only by IF is that this measure made review journals the most important, and this accorded well with Garfield's view of how the paradigms of science are defined and science advances.

**Citation Indexing and the Google Search Engine: Lawrence Page, Sergey Brin, and PageRank**

Garfield's purpose, method, and measures were integrated into the Web search engine developed by Lawrence Page and Segey Brin, who founded Google Inc. in the late 1990s while graduate students in computer science at Stanford University.  Google Inc. is a multinational corporation providing internet-related products and services such as searching, e-mail, cloud-computing, etc.  The basis of the company is a Web search engine.  A succinct and cogent explanation of how this search engine operates has been provided by Sherman (2005), who begins it by pointing out that "'searching the Web' with Google" is not an accurate statement, because, he writes, "What we're really searching is Google". (p. 4).  He describes Google as a massive Web harvesting machine that stores content and makes it searchable simultaneously by millions of people.  According to Sherman (2005, p. 7), like other search engines, Google's consists of three parts: a) the crawler—a.k.a. "spider" or "robot"—which is responsible for finding and harvesting all content discovered on the Web; b) the indexer, which is responsible for cataloging and storing information about Web content; and c) the query processor, which is the system you interact with to obtain search results.  In an explanation of the basics for webmasters Google (2012) likens searching the Web to "looking in a very large book with an impressive index telling you exactly where everything is located." The company states that, when you perform a Google search, its programs check its index to determine the most relevant search results to be returned ("served") to you.  Google divides the search process into three stages.  The first is called "crawling," whereby its crawler named "Googlebot" discovers new and updated pages to be added to the Google index.  To do this, the company utilizes a huge set of computers to crawl billions of Web pages with Googlebot that uses an algorithmic process to determine which sites to crawl, how often, and how many pages to fetch from each site. Googlebot's crawl process begins with a list of webpage URLs, generated from previous crawl processes, and, as Googlebot visits each of these websites it detects links on each page and adds them to its list of pages to crawl. New sites, changes to existing sites, and dead links are noted and used to update the Google index. The second stage is called indexing, whereby Googlebot processes each of the pages it crawls in order to compile a massive index of all the words it sees and their location on each page. The result is a huge network stored on a multitude of computers. The third and final stage is called "serving" by Google and occurs when a user enters a query and its machines search the index for matching pages to return the results believed the most relevant to the user. Google employs some 100 metrics to produce relevant document sets to search queries. Büttcher, Clarke, and Cormack (2010) define a document as "relevant", if "its contents (completely or partially) satisfy the information need represented by the query." Relevance is judged by human assessors and may be "*binary* ('relevant' or 'not relevant') or *graded* (e.g., 'perfect', 'excellent, 'good', 'fair', 'acceptable', 'not relevant', 'harmful'). Büttcher, Clarke, and Cormack state that the fundamental goal of relevance ranking is frequently expressed in terms of the "Probability Ranking Principle", which they phrase as follows: "*If an IR [Information Retrieval] system's response to each query is a ranking of the documents in the collection in order of decreasing probability of relevance, then the overall effectiveness of the system to its users will be maximized*" (p. 8).

The main determinant of relevance in the Google search engine is a citation or link analysis algorithm named PageRank. Garfield's ideas were incorporated into this algorithm not directly but as modified by the "influence" method developed by Computer Horizons, Inc. (CHI) during the 1970s. Computer Horizons Inc., whose president was Francis Narin (1976, p. ii), began the development of what it termed "evaluative bibliometrics" in July, 1970, under contract from the National Science Foundation. As stated by Narin (1976), the influence method had the following purpose:

> … [The] influence methodology…allows advanced publication and citation techniques to be applied to institutional aggregates of publications, such as those of departments, schools, programs, support agencies and countries, without performing an individual citation count. In essence, the influence procedure ascribes a weighted average set of properties to a collection of papers, such as the papers in a journal, rather than determining the citation rate for papers on an individual basis.

> The influence methodology is completely general, and can be applied to journals, subfields, fields, institutions or countries. (p. 183)

Here we will discuss the application of the CHI influence method to the assessment of U.S research-doctorate programs because of the possibility of gauging its relevance to human judgment.

In a key paper foreshadowing the basic concept underlying PageRank, Pinski and Narin (1976) set forth the basic idea behind the journal influence method that a citation from a highly prestigious journal should be given more weight than a citation from a less influential and obscure journal. They introduced the CHI method by first critiquing previously used measures of influence. According to them, total number of publications is only a measure of total activity, whereas total cites to a set of publications, while incorporating a measure of peer recognition, was size dependent with no meaning on the absolute scale. Most interesting and significant was their criticism of Garfield's impact factor, which, they noted, was size independent as the ratio of the number of citations the journal receives to the number of publications in a specified earlier time period but also has no meaning on the absolute scale. Pinski and Narin (1976) then listed three additional limitations of the impact factor, which they began thus:

> …Although the size of the journal, as reflected in the number of publications, is corrected for, the average length of individual papers appearing in the journal is not. Thus, journals which publish longer papers, namely review journals, tend to have higher impact factors. In fact the nine highest impact factors obtained by Garfield were for review journals. This measure can therefore not be used to establish a "pecking order" for journal prestige. (p. 298).

Pinski and Narin saw the superior citation rate of review articles as merely a function of their length, missing their important role in defining scientific paradigms. This is a common misevaluation of review articles. Thus, the creators of the "article influence score" now utilized as a measure in the recent *Journal Citation Reports* discount references from review articles, because "…a citation from a review article that has cursory references to large numbers of papers counts for less than a citation from a research article that cites only papers that are essentially related to its own argument" (Bergstrom, 2007, p. 314). The other two limitations attributed by Pinski and Narin (1976) to the impact factor were: 1) the citations were unweighted, with all citations being counted with equal weight, regardless of the citing journal; and 2) there was no normalization of the differing referencing characteristics of the various scientific fields in terms of number and time.

To counteract what they perceived to be the faults of Garfield's IF, Pinski and Narin (1976, p. 298) proposed the following three influences measures: (1) the influence weight of the journal, a size independent measure of the weighted number of citations a journal receives from the other journals, normalized by the number of references it gives to other journals; (2) the influence per publication for the journal, which is the weighted number of citations each article, note or review in a journal receives from other journals; (3) the total influence of the journal which is the influence per publication times the total number of publications. The influence weight of the journal was basically calculated by dividing the number of citations a journal received by the number of references it gave to other journals. Of the three influence measures, the second—the journal's influence per publication—was the most important, and Anderson, Narin, and McAllister (1978) verified this in their succinct summation of the purpose of the CHI method below:

> …The primary purpose of this methodology is to extract a measure of the influence
>
> of an individual journal from its citation relationship with other journals. In essence,
>
> the influence per paper for a journal is the weighted number of times an average paper
>
> in the journal is cited, where the weighting is based on the influence of the citing journal.
>
> Thus a citation from a highly prestigious journal such as the *Journal of Biological Chemistry*
>
> counts more heavily than a citation from a less prestigious journal. (p. 94)

The calculations presented by Pinski and Narin (1976) are complex, but their results can be easily demonstrated by two extreme examples p. 304). Thus, on one extreme, *Physical Review Letters* had a high influence weight (3.42), but being a letters journal, it had few references per publication (11.1). Its influence per publication—actually an enhanced IF—was 38.1, but it had 897 publications, which gave it a total influence of 34,186 (38.1*897). On the other extreme, *Reviews of Modern Physics*, a small review journal, had an influence weight 2.10 but had 116.9 references per publication. Its influence per publication was an extremely high 245.8, but since it only had 18 publications, its total influence was only 4,424 (18*245.8). In such a way Pinski and Narin reduced the importance of review journals to what they considered their true level.

Andersen, Narin, and McAllister (1978) tested the CHI influence method against peer ratings in the assessment of U.S. research-doctorate programs in mathematics and the sciences. Reputational ratings by peers had been the traditional method of evaluating U.S. graduate programs, and the evaluation of U.S. graduate programs conducted by Cartter (1966) under the auspices of American Council on Education in 1964 represented a milestone in that its survey questionnaire for obtaining such ratings was utilized by all future such studies. In the questionnaire, raters were asked to judge "the <u>quality of the graduate faculty</u>" (underlining in original), taking into consideration only their "scholarly competence and achievements," and to assign grades from 1 to 6 to the programs. In addition, the raters were given the option of not evaluating the programs by marking their questionnaire "Insufficient information" in the appropriate box. The Cartter assessment then assigned the grades the following numerical weights: Distinguished–5; Strong–4; Good–3; Adequate–2; Marginal–1; Not sufficient to provide acceptable doctoral training–0. These numerical weights were averaged to obtain a score for each program.  This method of obtaining and presenting peer ratings of the scholarly quality of university faculty was essentially replicated by the second American Council on Education evaluation of graduate education done in 1969 by Roose and Andersen (1970).  Andersen, Narin, and McAllister (1978) tested three bibliometric measures against the Roose-Andersen ratings in 10 fields: 1) total number of university papers in the field; 2) university influence per paper in the field; and 3) the total influence of the university's papers in the field, which was obtained by multiplying the total number of papers by the average influence per paper.  The bibliometric ratings are based on 127,000 university papers, from 450 journals in the 10 fields from 1965 to 1973.  The Roose-Andersen peer ratings were found to correlate most highly with the total influence of the university's papers, followed closely by correlations with the total number of papers, and much less closely with the average influence per paper. A partial correlation and regression analysis indicated that the Roose-Andersen ratings had two additive components: bibliometric size and bibliometric quality.  The CHI influence method was also employed for the mathematical and physical sciences in the 1981 assessment of U.S. research-doctorate programs done under the auspices of the American Council of Learned Societies, American Council on Education, National Research Council, and Social Science Research Council (Jones, Lindzey, and Coggeshall, 1982, pp. 15 and 27-29), which used the same method of peer ratings as did Cartter and Roose-Andersen.  The 1981 assessment employed two publication measures.  The first was the number of published articles attributed to the program, which was compiled by CHI from the 1978 and 1979 *Science Citation Index*, whereas the second was the estimated "overall influence" of these articles, which was a product of the number of articles attributed to the program times estimated influence of the journals in which these articles appeared.  The influence of a journal was determined by the weighted number of times, on the average, an article in that journal was cited—with references from frequently cited journals counting more heavily.  Of the six fields being assessed, the correlation of the peer ratings was higher with the overall influence of the articles than with the number of articles in five, rising in the following manner (p.

166): chemistry—from 0.80 to 0.86; computer sciences—from 0.70 to 0.77; geosciences—from 0.75 to 0.77; mathematics—from 0.75 to 0.83; and physics—from 0.85 to 0.86. Only in statistics/biostatistics did the correlation fall from 0.70 to 0.67. From the perspective of the development of the Google PageRank, the lesson to be drawn from the above CHI findings is that links from documents, to which there are many links, produce results more in accord with subjective human judgment than links from documents, to which there are few links.

The CHI influence method was integrated into search engine technology by Jon M. Kleinberg, an assistant professor of computer science at Cornell University, who was in communication with Page and Brin during the critical early period, when they were developing their own search engine and establishing their company. A summary of Kleinberg's ideas as well as an interesting insight into his relationship with Page and Brin is provided by Battelle (2005, pp. 70-71 and 81-82) in his history of Google Inc. Battelle interviewed Kleinberg in writing this book. Kleinberg came into close contact with Page and Brin while a member of the Clever Project at the IBM Almaden Research Center in San Jose, California. The purpose of the Clever Project was to develop improved techniques for finding information in the clutter of cyberspace by exploiting hyperlink structure of the WWW, and the proximity of San Jose to Palo Alto enabled Kleinberg to visit the Stanford campus to compare notes with Page and Brin. In a feature article in *Scientific American* Members of the Clever Project (1999) wrote a synopsis of their work. This article not only described the work of the Clever Project, linking it to citation analysis through Garfield's impact factor and CHI's modification of it, but also referred to similar work being done on Google by Brin and Page at Stanford. Narin himself took note of the *Scientific American* article, and in a contribution to a festschrift dedicated to Garfield on the development of science indicators in the U.S., Narin, Hamilton, and Olivastro (2000, p. 340) briefly commented upon the similarity of the concept underlying the Clever Project's work to the CHI influence method without indicating any awareness of the significance of what was taking place.

As its name implies, the purpose of the Clever Project was the development of an experimental Web search engine, which was code-named Clever. The research of this project was summarized in a simplified form by Members of the Clever Project (1999) in the *Scientific American* article above and in a more authoritative version by Kleinberg (1999) in the *Journal of the ACM*. Members of the Clever Project (1999) succinctly posed the problem, which the project tried to solve, thus:

> Every day the World Wide Web grows by roughly a million electronic pages, adding to
>
> the hundreds of millions already on-line. This staggering volume of information is loosely
>
> held together by more than a billion annotated connections, called hyperlinks….

> But because of the Web's rapid, chaotic growth, the resulting network of information lacks organization and structure. In fact, the Web has evolved into a global mess of previously unimagined proportions. Web pages can be written in any language, dialect or style by individuals with any background, education, culture, interest and motivation. Each page might range from a few characters to a few hundred thousand, containing truth, falsehood, wisdom, propaganda or sheer nonsense. How, then, can one extract from this digital morass high-quality, relevant pages in response to a specific need for certain information?   (p. 54)

To do this, according to the Members, people utilize search engines, which, at their most basic level, maintain lists, collectively known as an index, for every word, of all known Web pages containing that word.  They note that determining what information to return in response to user queries remains daunting, because, for example, certain search terms appear on millions of Web pages, resulting in what Kleinberg (1999) called the "*Abundance Problem: The number of pages that could reasonably be returned as relevant is far too large for a human user to digest*" (p. 606).  One difficulty, the Members wrote, is there is no exact and mathematically precise measure of what is "best," which lies in the eyes of the beholder, and certain search engines use heuristics to rank or prioritize pages.  Simple heuristics like ranking pages by the number of times they contain the search terms or instances these appear earlier in the text were rejected by the Members for various reasons—a major one being that human language is awash in synonymy and polysemy.   After considering text-based approaches, Kleinberg declared, "Indeed, one suspects that there is no purely *endogenous* measure of the page that would allow one to properly assess its authority" (p. 606).

Seeing no intrinsic bases in the content of Web pages for defining relevant document sets in response to search queries, the Clever Project switched the focus to the analysis of the hyperlink structure connecting these pages.  In discussing this structure we will use the terminology recommended by Vaughan (2005, p. 949) for types of links.  According to this terminology, an "inlink,"—a.k.a., "backlink"—is a link coming into a Web site, document, or page.  In citation terms it is the equivalent of a "citation."  On the other hand, an "outlink"—a.k.a., "outgoing link"—is a link going from the Web site, document, or page.  In citation terms it is the equivalent of a "reference."  Then there are "external links" (links coming from outside the item being linked) and "internal links" (links coming from within the item being linked).  The latter can be thought of as "self-citations."  Underlying the Clever Project's focus on the Web hyperlink structure was the concept of "authority," and Kleinberg (1999) explained this concept and the role of hyperlinks in defining it thus:

> This notion of authority, relative to a broad-topic query, serves as a central focus in our work. One of the fundamental obstacles we face in addressing this issue is that of accurately modeling authority in the context of a particular query topic. Given a particular page, how do we tell whether it is authoritative?
>
> …Hyperlinks encode a considerable amount of latent human judgment, and we claim that this type of judgment is precisely what is needed to formulate a notion of authority. Specifically, the creation of a link on the www represents a concrete indication of the following type of judgment: the creator of page $p$, by including a link to page $q$, has in some measure *conferred authority* on $q$.  (p. 606)

Kleinberg noted that links opened the opportunity to find potential authorities purely through the pages that point to them, and this caused the Clever Project to base its search engine on a system of "authorities" and hubs, which the Members (2009) succinctly described thus:

> In addition to expert sites that have garnered many recommendations, the Web is full of another type of page: hubs that link to those prestigious locations, tacitly radiating influence outward to them. Hubs appear in guises ranging from professionally assembled lists on commercial sites to inventories of "My Favorite Links" on personal home pages. So even if we find it difficult to define "authorities" and "hubs" in isolation, we can state this much: a respected authority is a page that is referred to by many good hubs; a useful hub is a location that points to many valuable authorities.  (p. 58)

Explained simply, the Clever search engine was designed to operate in the following manner (Kleinberg, 1999, pp. 607-617; Members, 1999, pp.57-58).  It would first assemble a small set of some 200 pages on a given topic with a standard text-based search engine such as AltaVista.  Then, through an iterative process of hyperlink analysis, it would expand this set and determine the best "authorities" and "hubs" in this set being expanded by tracing links, using two measures—out-degree (number of nodes a given node has links to) and in-degree (the number of nodes that have links to the given node). The pages were scored in a manner that a page that had more high-scoring hubs pointing to it had a higher authority score, whereas a page that pointed to more high-scoring authorities garnered a higher hub score. This iterative process was continued until equilibrium was reached.

Both the Members (1999, pp. 58-60) and Kleinberg (1999, pp. 617-619) discussed the connections of the Clever algorithm to citation analysis. The discussion of the Members was brief, mentioning Garfield's impact factor, the *SCI*, and the CHI influence method, but—as seen above—it was enough to catch the eye of Narin himself. That of Kleinberg was theoretically at a much higher level, and he began by placing bibliometrics and citations within the context of social network analysis or the analysis of the articulated patterns of connections in the social relations of individuals, groups, or other collectivities. Kleinberg noted that the study of social networks had developed several ways to measure "the relative *standing*—roughly, "importance"— of individuals in an implicitly defined network" (p. 617), and he then connected citations to the Clever Project thus:

> …Research in bibliometrics has long been concerned with the use of citations to produce quantitative estimates of the importance and "impact" of individual scientific papers and journals, analogues of our notion of authority. In this sense, they are concerned with evaluating standing in a particular type of social network—that of papers or journals linked by citations. (p. 618)

Noting that the best-known measure in the field was Garfield's impact factor, he defined it in hypertext terms by stating that "the impact factor is a ranking measure based fundamentally on a pure counting of the in-degrees of nodes in the network" (p. 618). He then integrated the CHI influence method into the Clever algorithm in the following manner:

> Pinski and Narin [1976] proposed a more subtle citation-based measure of standing, stemming from the observation that not all citations are equally important. They argued that a journal is "influential" if, recursively, it is heavily cited by other influential journals. One can recognize a natural parallel between this and our self-referential construction of hubs and authorities…. (p. 618)

Kleinberg (1999) then set forth distinctions between bibliomentrics and Webology with potentially important implications thus:

…[T]he World Wide Web and the scientific literature are governed by very different principles, and this contrast is nicely captured in the distinction between Pinski–Narin influence weights and the hub/authority weights that we compute. Journals in the scientific literature have, to a first approximation, a common purpose, and traditions such as the peer review process typically ensure that highly authoritative journals on a common topic reference one another extensively. Thus, it makes sense to consider a one-level model in which authorities directly endorse other authorities. The www, on the other hand, is much more heterogeneous, with www pages serving many different functions…. Moreover, for a wide range of topics, the strongest authorities consciously do not link to one another…. Thus, they can only be connected by an intermediate layer of relatively anonymous hub pages, which link in a correlated way to a thematically related set of authorities….. (p. 619).

Battelle (2005) traces the course of Kleinberg's argumentation here in the following interesting manner:

> …Kleinberg first defines a term (biblometrics). He then cites the authority in the space (the legendary Eugene Garfield, who is widely credited as the father of citation analysis), and proceeds to cite those who have built upon Garfield's work [Pinski and Narin]. Finally, Kleinberg puts forward his own conclusions, based on his theories of hubs and authorities. (pp. 70-71)

One conclusion, which Kleinberg (2009, p. 629) drew, was that measures of impact and influence in bibliometrics arguably did not require an analogous formulation of the role played by the hubs. As will be seen, this conclusion has implications for Google Scholar, whose coverage is focused on academic literature.

While working on the Clever Project, Kleinberg heard about the work of Page and Brin at Stanford on a search engine then called "BackRub" and in the summer of 1997 visited the campus to compare notes. Kleinberg had just finished an early version of his seminal paper on authoritative Web sources, and he encouraged Page to publish an academic paper on his search engine. As told by Kleinberg to Battelle (2005, pp. 81-82), during their conversation Page manifested reluctance, because he feared that somebody might steal his ideas. In Kleinberg's words, "[Page] felt like he had the secret formula," and "It did seem a bit strange at the time" (Battelle, 2005, 82). However, by the end of the conversation Page and Kleinberg agreed to cite each other in their papers.

Shortly after this conversation, Page and Brin changed the name of their search engine to Google and established their company. In 1998, the year in which Google was incorporated, Page and Brin set forth the principles, on which their search engine operated, in two critical documents: a Stanford technical report by Page et al. (1998); and a paper delivered to the seventh International WWW Conference by Brin and Page (1998). Later Page (2001) summarized these principles in a patent application. The conference paper is of interest, for it is here that Brin and Page (2008) stated that they chose the name Google, "because it is a common spelling of googol, or $10^{100}$ and fits well with our goal of building very large-scale search engines" (p. 108). Page and Brin did not fully reveal the intellectual origins of their thought in their two 1998 papers. While Kleinberg was cited in both of these documents, only a minor, rather uninformative paper by Garfield (1995) was referenced by Page, L. et al. (1998). However, in his patent application Page did reveal these origins, citing not only Kleinberg but also the *Science* article by Garfield (1972), where he summarized the findings of the ISI project leading to the *JCR*s and introduced the impact factor to the broad scientific community, as well as the article by Pinski and Narin (1976), which set forth the CHI influence method. It will be seen from what follows that the chief Google measure of authoritativeness—PageRank—was at its basis a further development of the concepts and techniques originally developed in citation analysis and their application to a higher level of technology.

PageRank was first introduced in the Stanford technical report entitled "The PageRank Citation Ranking: Bringing Order to the Web" by Page et al. (1998). Here PageRank is introduced as "a method for rating Web pages objectively and mechanically, effectively measuring the human interest and attention devoted to them" (p. 1). Like Kleinberg, Page et al. found the solution to the problems posed for information retrieval by the complexity and heterogeneity of the Web not in the text of the web pages but in the link structure of the Web, which can be utilized "to produce a global 'importance' ranking of every web page" (p. 1) Although never stated by Kleinberg, the ability to define relevancy in terms not of words but of linkages is the central premise of citation indexing, which—as seen above—Garfield (1955) set forth in his *Science* article introducing this concept, where he rejected conventional alphabetical and subject indexes as inadequate and described the linkage-based citation index as "an association-of-ideas index" (p. 108). Page et al. (1998) began their analysis how to bring order to the Web by defining the differences between web pages and academic publications, on which citation analysis was based, thus:

> …there are a number of significant differences between web pages and academic
> publications. Unlike academic papers which are scrupulously reviewed, web pages

> proliferate free of quality control or publishing costs… Further, academic papers are well defined units of work, roughly similar in quality and number of citations, as well as in their purpose to extend the body of knowledge. Web pages vary on a much wider scale than academic papers in quality, usage, citations, and length… [T]he simplicity of creating and publishing web pages results in a large fraction of low quality web pages that users are unlikely to read. (pp. 1-2)

Despite these significant differences, Page et al. (1998) nevertheless declared somewhat puzzlingly:

> It is obvious to try to apply standard citation analysis techniques to the web's hypertextual citation structure. One can simply think of every link as being like an academic citation. So, a major page like http://www.yahoo.com/ will have tens of thousands of backlinks (or citations) pointing to it. (p. 2)

While noting that the high number of inlinks to the Yahoo home page implied its importance, Page et al. (1998) rejected simple inlink counts as a method of measuring Web pages due to characteristics of the Web not present in normal academic citation databases. Instead they proposed PageRank, explaining it and justifying it by reference to the distributional structure of the Web thus:

> Web pages vary greatly in terms of the number of backlinks they have. For example, The Netscape home page has 62,804 backlinks in our current database compared to most pages which have just a few backlinks. Generally, highly linked pages are more "important" than pages with few links. Simple citation counting has been used to speculate on the future winners of the Nobel Prize…. PageRank provides a more sophisticated method for doing citation counting.
>
> The reason that PageRank is interesting is that there are many cases where simple citation counting does not correspond to our common sense notion of importance. For example, if a web page has a link to the Yahoo home page, it may be just one link but it is a very important one. This page should be ranked higher than many pages with more links but from obscure places. PageRank is an attempt to see how good an approximation to \importance" can be obtained just from the link structure. (p. 3)

The influence of the CHI influence method on Page and Brin is quite obvious here.

PageRank is extraordinarily complex both logically and mathematically but is explained cogently by Page (2001, pp. 2-3) in layperson terms in his patent application. According to Page, the measure provides a method that is scalable and can be applied to large databases such as the WWW. He writes that PageRank takes advantage of the linked structure of the Web to assign a rank to each document in the database, where document rank is a measure of the importance of the document. Rather than determining relevance only from the intrinsic content of a document, he states, PageRank determines importance from the extrinsic relations between documents, stating, "Intuitively, a document should be important (regardless of its content) if it is highly cited by other documents" (p. 2). However, according to him, not all citations are equal, and a citation from an important document is more important that a citation from a relatively unimportant document. Thus, he reasons, the importance of a page, and hence its rank, should depend not just on the number of its citations but on the importance of the citing documents as well. This, he states, implies a recursive definition of rank: the rank of a document is a function of the ranks of the documents which cite it. PageRank calculates document ranks by an iterative procedure on a linked database. Switching the basis of the argument to probability, Page writes that the importance of a page is directly related to the steady-state probability that a random Web surfer ends at the page after following a large number links. According to him, there is a larger probability that the surfer will wind up at an important page than at an unimportant page.

In the patent application Page (2001) stated that "a high rank indicates that a document is considered valuable by many people or by important people" (p. 3), and, in their conference presentation, Brin and Page (1998) described PageRank as "an objective measure of …citation importance that corresponds well with people's subjective idea of importance" and, because of this correspondence, "an excellent way to prioritize the results of Web keyword searches" (p. 109). However, these are merely assertions, and Page and Brin presented little evidence of how well PageRank results correspond with human subjective judgment. That which they did had a considerable component of what Kleinberg (1999) once termed *res ipsa loquitur* ("the thing speaks for itself") (p. 627) in reference to evaluations of the efficacy of his Clever engine. An interesting piece of evidence presented by Page et al. (1998, p. 9) was a comparison of what was retrieved in an informal test in response to the query "University" by Google based on PageRank and Altavista based on text. Google yielded a list of top universities, whereas Altavista returned random looking web pages matching the query "University." From this it does seem that PageRank delivers results more in conformance with human subjective judgment just as (see above, pp. 17-18) publication counts weighted by the CHI influence method more conformed with peer ratings of research-doctorate programs than did un-weighted counts.

 Google does not only rely on the extrinsic relationships between documents to form its relevant sets but also uses what Sherman (2005) calls "'on-the-page' metrics" (p.12), which utilize the terms of a search query. He notes that Google likes pages where all the search terms appear, likes better the pages where the search terms are near each other, and likes best the pages where search terms appear in the order they are input, as in a phrase. Other on-the-page metrics noted by Sherman are: search terms appearing in the title of the page; search terms appearing in unique font such as bold or italic; and the frequency of the appearance of the search terms on the page. In the longer version of their conference paper available on the Web, Brin and Page (1998) discuss these on-the-page ranking methods, giving the following description of on-the-page ranking of a multi-word query:

> For a multi-word search, the situation is more complicated. Now multiple hit lists must be scanned through at once so that hits occurring close together in a document are weighted higher than hits occurring far apart. The hits from the multiple hit lists are matched up so that nearby hits are matched together. For every matched set of hits, a proximity is computed. The proximity is based on how far apart the hits are in the document (or anchor) but is classified into 10 different value "bins" ranging from a phrase match to "not even close". Counts are computed not only for every type of hit but for every type and proximity. Every type and proximity pair has a type-prox-weight. (Article FP11)

Given the topic of this paper, it should be pointed out that this method of determining relevance can be a serious source of error with search inquiries involving the names of scholars and scientists doing collaborative research. At the lower ranks, where the probability of relevance is rather low and the probability of error is rather high, Google has a tendency to combine the first name of one collaborator with the last name of another and include the document in the retrieved set.

An extremely informative insight into the operation of the Google search engine was provided by Kleinberg (1999), who compared the method adopted by Page and Brin to the one he adopted for Clever on the following two points thus:

> One of the main contrasts between our approach and the page-rank methodology is that—like Pinski and Narin's formulation of influence weights—the latter is based on a model in which authority is passed directly from authorities to other authorities, without interposing a notion of hub pages. Brin and Page's use of random jumps to uniformly selected pages is a way of dealing with the resulting problem that many authorities are essentially "dead-ends" in their conferral process.
>
> It is also worth noting a basic contrast in the application of these approaches to www search. In Brin and Page…, the page-rank algorithm is applied to compute ranks for all the nodes in a 24-million-page index of the www; these ranks are then used to order the results of subsequent text-based searches. Our use of hubs and authorities, on the other hand, proceeds without direct access to a www index; in response to a query, our algorithm first invokes a text-based search and then computes numerical scores for the pages in a relatively small subgraph constructed from the initial search results.

The Members of the Clever Project (1999, p. 60) note that the elimination of the preliminary step of forming textual sets enables Google to respond much more quickly to queries. Since we are dealing with Google Scholar, which covers academic literature like the citation indexes, it should be pointed out Kleinberg (1999, p. 629) argued that scientific literature is much different from the WWW and that therefore measures of impact and influence in bibliometrics have typically neither had nor needed an analogous formulation of the role hubs play in conferring authority.

**The Validation of Google Scholar (GS)**

It has been shown above that Garfield's theory of citation indexing and the Google search engine are based upon the same theoretical premises. These premises were first set forth by Garfield (1955) in his *Science* article entitled "Citation Indexes for Science."  Simply stated, these premises posit that coherent subject sets are better defined by linkages—i.e., citations (inlinks or backlinks)—than by words. Narin and the CHI influence method further developed these premises by positing that citations from documents that themselves are more highly cited are better able to construct sets that conform to human subjective judgment. However, as a result of his Law of Concentration, Garfield built this principle into the *Science Citation Index* by limiting its coverage as far as possible to the most highly cited titles forming multidisciplinary core of the scientific journal system. The Google search engine incorporates these premises into its PageRank algorithm, which weights more heavily citations from documents that themselves are more heavily cited. Given the similarity of their premises, Google Scholar (GS) should yield results that are in conformance with Garfield's findings through citation analysis. This is the hypothesis which will be tested in this paper, and it will be done by stealing a page from Garfield's book by analyzing GS citations to the works of Nobel laureates in chemistry—the discipline in which Garfield did his pioneering work.

To test the hypothesis, citations to the works of 5 Nobel laureates in chemistry were downloaded from Google Scholar with Anne-Wil Harzing's revolutionary Publish-or-Perish (PoP) software that has established statistical control over this database. In order to understand the test, it is necessary to have knowledge of GS coverage and the capabilities of Harzing's program. Concerning the first, Google (2011b) gives fairly complete description of GS coverage in the help section to that database. Here it states:

> Google Scholar includes journal and conference papers, theses and dissertations, academic books, pre-prints, abstracts, technical reports and other scholarly literature from all broad areas of research. You'll find works from a wide variety of academic publishers, professional societies and university repositories, as well as scholarly articles available anywhere across the web.  Google Scholar also includes court opinions and patents.

Google qualifies this by stating that, although it attempts to be comprehensive, it cannot guarantee uninterrupted coverage of any particular source and that it is necessary to remove websites, when they become unavailable to its search robots or to a large number of Web users.  It emphasizes that GS indexes, not journals, but academic papers and that shorter articles, such as book reviews, news sections, editorials, announcements and letters, may or may not be included.  Google greatly strengthened GS coverage by including in it Google Book records that allows the full-text searching of millions of books supplied by both publishers and libraries (Google, 2011a; Baksik, 2011).

The data, which was utilized to test the above hypothesis, was downloaded from Google Scholar (GS) in September, 2011, by Anne-Wil Harzing (2010) with her revolutionary Publish or Perish (PoP) computer program, which is freely available on her Web site at http://www.harzing.com, and was given to the author of this paper.  Utilization of GS to evaluate scientific and scholarly research is highly controversial, and much of the criticism is directed against the relevance and quality of the elements of the sets formed by GS in response to search queries.  A major fault of GS is that by itself it does not provide a mechanism for easily retrieving and forming datasets suitable for statistical analyses.  Harzing's PoP program was specifically designed to overcome this fault.  It is a software program that can retrieve and analyze raw GS citations, form relevant sets from these citations, calculate a wide range of metrics based on these sets, and export the data to Excel for further analysis.  The most articulate critic of utilizing GS to evaluate research, Peter Jacsó (2009), tested Harzing's software and found it to be "a swift and elegant tool to provide the essential output features that Google Scholar does not offer" (p. 1189).

The data (Harzing, 2013; Harzing, Forthcoming), consists of GS citations to the works of five chemistry laureates, and a primary consideration in their selection was the year their prizes were won:

1990, 2000, 2008, 2009, and 2010. This yielded a selection of three prizes one decade apart—1990, 2000, and 2010—and three prizes one year apart—2008, 2009, and 2010—and it was done to capture the influence of time. In years with multiple winners for a particular field Harzing selected the first Nobelist unless this laureate had a particularly common name that would complicate data retrieval. For each Nobelist, Harzing (2013, pp. 1064-1065) constructed two inlinks measures. The first was total inlinks, which represented the most comprehensive measure of the laureate's impact. The second was his h-index, which she considered "the best indication of the number of publications that had achieved a significant impact" (p. 1065). PoP automatically calculates the h-index. The h-index has become one of the leading measures for the evaluation of scientists. It was created by J. E. Hirsch, a physicist at the University of California at San Diego. In his initial formulation Hirsch (2005) defined his h-index thus: "A scientist has index h if h of his or her Np papers have at least h citations each and the other (Np - h) papers have ≤ h citations each" (p. 16569), where Np is the number of papers published over n years. Hirsch (2007) in his next paper on this measure modified his initial formulation in the following important way: "The h index of a researcher is the number of papers **coauthored** [emphasis added] by the researcher with at least h citations each" (p. 19193). Having made this important modification, Hirsch (2010) openly acknowledged in a following paper that "the most important shortcoming of the h-index is that it does not take into account in any way the number of coauthors of each paper," stating, "This can lead to serious distortions in comparing individuals with very different coauthorship patterns…." (p. 742). This weakness particularly affects a field like chemistry, which is a highly collaborative field, and GS retrieves works attributable to a scientist no matter what the position of this scientist in the authorship structure of the work—be it primary author, secondary author, or even editor, as in the case of books.

Harzing paid special attention to publications included in the h-index, verifying them individually to ensure they were published by the Nobelist in question and merging any publications with substantial

| TABLE 1.  GOOGLE SCHOLAR (GS) H-INDEX PUBLICATIONS OF NOBEL PRIZE WINNERS IN CHEMISTRY ||||||||
| Prize Winner || Quantitative Characteristics of H-Index Publications |||| Bibliographic Type |||
| Name | Year | H-Index | Maximum Cites | Cites Range | Total Cites | Journal Articles | Books | Conference Proceeding |
| Elias J. Corey | 1990 | 97 | 1225 | 1129 | 23216 | 97 | 0 | 0 |
| Alan J. Heeger | 2000 | 123 | 3321 | 3199 | 49266 | 123 | 0 | 0 |
| Osamu Shimomura | 2008 | 50 | 929 | 878 | 5852 | 49 | 1 | 0 |
| Ada E. Yonath | 2009 | 36 | 637 | 602 | 4381 | 36 | 0 | 0 |
| Ei-ichi Negishi | 2010 | 49 | 797 | 748 | 6692 | 48 | 1 | 0 |

stray records, if they were on the h-index threshold.  In dealing with WWW-type distributions, the h-index has a major advantage in that such distributions tend to become incoherent with incorrect and irrelevant elements in the lower range, the h-index is robust against that.  Truncating the distribution at h-index removes a lot of what can be termed "clutter" from the set.

Table 1 above lists the chemistry laureates selected for analysis, the year or their prize, the quantitative characteristics of their h-index publications in terms of GS cites, as well as the bibliographic characteristics of their h-index publications.  Here the effect of time is clearly visible.  Thus, two earliest laureates—Corey (1990) and Heeger (2000)—have h-indexes from 97 to 123, maximum GS cites from 1225 to 3321, GS cites ranges from 1129 to 3199, and total GS cites from 23216 to 49266.  The corresponding measures for the three later laureates—Shimomura (2008), Yonath (2009), and Negishi (2010)—are much lower.  Thus, their h-indexes are from 36 to 50, their maximum GS cites from 637 to 929, their GS cites range from 602 to 878, and their total GS cites from 4381 to 6692.  In contrast, the bibliographic characteristics of their h-index works manifest an almost stunning uniformity.  Almost all were journal articles except for two books.  However, as will be seen, these books reinforce Garfield's finding through citation analysis.

Tables 2A, 2B, and 2C below present data, which tests how well the laureates' h-index publications defined in terms of GS citations conform to Garfield's Law of Concentration.  To construct these tables, stratified random samples of every third journal article starting from the highest in GS cites were selected.  It was then determined whether these article were published in journals

**TABLE 2A.** *SCIENCE CITATION INDEX JOURNAL CITATION REPORT* (SCI JCR) RANKINGS OF JOURNALS PUBLISHING PAPERS AUTHORED BY NOBEL PRIZE WINNERS IN YEAR OF APPEARANCE: NUMBER/PERCENT INDEXED AND TOTAL CITES RANK VS. IMPACT FACTOR RANK.

| Prize Winner | | | Number Indexed | | | Total Cites Rank vs. Impact Factor Rank | |
|---|---|---|---|---|---|---|---|
| Name | Year | Sample Size | Indexed | Not Indexed | % Indexed | No. Higher by TC Than IF | %. Higher by TC Than IF |
| Elias J. Corey | 1990 | 30 | 30 | 0 | 100.0% | 28 | 93.3% |
| Alan J. Heeger | 2000 | 41 | 41 | 0 | 100.0% | 34 | 82.9% |
| Osamu Shimomura | 2008 | 17 | 17 | 0 | 100.0% | 15 | 88.2% |
| Ada E. Yonath | 2009 | 12 | 11 | 1 | 91.7% | 7 | 63.6% |
| Ei-ichi Negishi | 2010 | 16 | 16 | 0 | 100.0% | 15 | 93.8% |

**TABLE 2B.** *SCIENCE CITATION INDEX JOURNAL CITATION REPORT* (SCI JCR) RANKINGS OF JOURNALS PUBLISHING PAPERS AUTHORED BY NOBEL PRIZE WINNERS IN YEAR OF APPEARANCE: RANKINGS BY TOTAL CITES

| Prize Winner | | | Total Cites Ranks of Those Indexed | | | | | | |
|---|---|---|---|---|---|---|---|---|---|
| Name | Year | Sample Size | No Ranked in Top 500 | % Ranked in Top 500 | No. Ranked 501 to 1000 | % Ranked 501 to 1000 | No. Ranked Below 1000 | %. Ranked Below 1000 | Median Rank |
| Elias J. Corey | 1990 | 30 | 30 | 100.0% | 0 | 0.0% | 0 | 0.0% | 22.5 |
| Alan J. Heeger | 2000 | 41 | 38 | 92.7% | 0 | 0.0% | 3 | 7.3% | 22 |
| Osamu Shimomura | 2008 | 17 | 13 | 76.5% | 2 | 11.8% | 2 | 11.8% | 159 |
| Ada E. Yonath | 2009 | 12 | 9 | 81.8% | 2 | 18.2% | 0 | 0.0% | 22 |
| Ei-ichi Negishi | 2010 | 16 | 16 | 100.0% | 0 | 0.0% | 0 | 0.0% | 13.5 |

indexed by the *Science Citation Index* (*SCI*) in the year they were published and, if so, what was the journal's *SCI Journal Citations Report* rank in that year by both total cites (TC) and impact factor (IF)—whether in the top 500, from 501 to 1000, or below the rank of 1000. The results were not only a vindication of Garfield's theory of citation indexing but also conversely of Google Scholar and the PageRank algorithm. Table 2A shows that, of the sample, all the laureates h-index articles except one by

| TABLE 2C. *SCIENCE CITATION INDEX JOURNAL CITATION REPORT* (SCI JCR) RANKINGS OF JOURNALS PUBLISHING PAPERS AUTHORED BY NOBEL PRIZE WINNERS IN YEAR OF APPEARANCE: RANKINGS BY IMPACT FACTOR ||||||||||
|---|---|---|---|---|---|---|---|---|---|
| Prize Winner | | | Impact Factor Ranks of Those Indexed |||||||
| Name | Year | Sample Size | No Ranked in Top 500 | % Ranked in Top 500 | No. Ranked 501 to 1000 | % Ranked 501 to 1000 | No. Ranked Below 1000 | %. Ranked Below 1000 | Median Rank |
| Elias J. Corey | 1990 | 30 | 29 | 96.7% | 1 | 3.3% | 0 | 0.0% | 131.5 |
| Alan J. Heeger | 2000 | 41 | 33 | 80.5% | 2 | 4.9% | 6 | 14.6% | 202 |
| Osamu Shimomura | 2008 | 17 | 11 | 64.7% | 1 | 5.9% | 5 | 29.4% | 200 |
| Ada E. Yonath | 2009 | 12 | 9 | 81.8% | 1 | 9.1% | 1 | 9.1% | 67 |
| Ei-ichi Negishi | 2010 | 16 | 13 | 81.3% | 3 | 18.8% | 0 | 0.0% | 136 |

Yonath were published in journals covered by the SCI in the year they were published and that these journals' TC rank was generally higher than their IF rank from a low of 63.6% to a high of 93.8% of the times. Table 2B analyzes the TC rankings of the journals publishing the laureates' articles, and it shows that, of the sample, most of the articles were published in journals in the top 500 by TC rank—from a low of 76.5% to a high of 100.0%--with a median rank for the top 4 laureates of 13.5 to 22.5. Here Shimomura (2008) is an outlier with a median rank of 159—low but still in the top 500. Table 2C above presents data on the IF rankings the journals in which the laureates' articles were published at the time of publication, and here, again the laureates tended to publish in the top 500 by IF rank but at a lower rate and level. Thus, of the sample, the percentage of articles in journals in the top 500 by IF rank ranged from 64.7% to 96.7% with median ranks from 67 to 136 lower generally than those of TC medians except for Shimomura but still in the top 500 by IF rank.

| TABLE 3. AUTHORSHIP PATTERN OF WORKS AUTHORED BY NOBEL PRIZE WINNERS ||||||||||||
| Prize Winner || | No. Authors || Position of Nobel Prize Winner ||||||||
| Name | Year | Sample Size | Range | Median | Primary | % Primary | 2nd to 5th | % 2nd to 5th | 6th to 10th | % 6th to 10th | 11th and Lower | % 11th and Lower |
|---|---|---|---|---|---|---|---|---|---|---|---|---|
| Elias J. Corey | 1990 | 33 | 1 to 8 | 2 | 25 | 75.8% | 7 | 21.2% | 1 | 3.0% | 0 | 0.0% |
| Alan J. Heeger | 2000 | 41 | 1 to 6 | 4 | 2 | 4.9% | 36 | 87.8% | 3 | 7.3% | 0 | 0.0% |
| Osamu Shimomura | 2008 | 17 | 1 to 10 | 5 | 7 | 41.2% | 7 | 41.2% | 3 | 17.6% | 0 | 0.0% |
| Ada E. Yonath | 2009 | 12 | 1 to 20 | 7 | 4 | 33.3% | 2 | 16.7% | 3 | 25.0% | 3 | 25.0% |
| Ei-ichi Negishi | 2010 | 16 | 1 to 7 | 2.5 | 10 | 62.5% | 5 | 31.3% | 1 | 6.3% | 0 | 0.0% |
| *PERCENT TIMES LAUREATES WERE PRIMARY AUTHOR OF THE WORKS = 40.3%* ||||||||||||

Table 3 above presents data on the authorship structure of the GS citation h-index publications of the chemistry laureates. This is important because GS retrieves documents of researchers no matter what their position in the authorship structure. Once again a stratified random sample of every third publication was taken, starting from the publication highest in GS cites. The number of authors per publication ranged from 1 - 6 for Heeger (2000) to 1 – 20 for Yonath (2009), with the median number of authors ranging from 2 for Corey (1990) and 7 for Yonath (2009). The percentage of times the laureate was primary author ranged from a low of 33.3% for Yonath (2009) to a high of 75.8% for Corey (1990). Overall the percentage of times the laureates were primary authors was 40.3%. Heeger (2000) is an interesting case. With a range of 1-6 authors and a median number of authors of 4, he was primary author only twice (4.9%), $2^{nd}$ -$5^{th}$ author 36 times (87.8%), and even last or $6^{th}$ author 3 times (7.3%). Yonath (2009) is the obvious outlier here with the highest authorship range (1-20), highest median number of authors (7), and her authorship positions evenly spaced across the entire authorship range, being the $11^{th}$ author or lower 25.0% of the time. This could be the result of her being the only female laureate in the sample. All in all, the data indicates that it is extremely difficult—if not impossible—to identify the most important contributor by authorship position alone.

For comparative purposes, the position of the Nobel laureates in the authorship structure of review articles was analyzed. The data is presented in Table 4 below, and it is most revealing. In general, the number of authors in review articles is much smaller—ranging from 1 to a median of 3—and the laureates are the primary authors a larger percentage of the time than in the non-review articles—68.8% vs. 40.3%. Negishi (2010) was most productive in review articles—6—and he was the primary author 100.0% of the times. From this it can be seen that the Nobelists play an important role in the formulation of the paradigms of their field. This conclusion is reinforced by the nature of the two books produced by the laureates. One was by Osamu Shimomura and entitled *Bioluminescence: Chemical Principles and Methods* (Hackensack, N.J.: World Scientific, 2006. 470 pages), and it is described as providing "a comprehensive overview of the biochemical aspects of the luminous organisms currently known" and

| TABLE 4. GOOGLE SCHOLAR (GS) CITES RANKS AND AUTHORSHIP PATTERNS OF REVIEW ARTICLES BY LAUREATES ||||||||||||
| Prize Winner ||| Review Articles || | Authorship Position of Nobel Prize Winner |||||||
| Name | Year | H-Index | Number Review Articles | GS Cites Rank Position or Median | Author Number or Median | Primary | % Primary | 2nd to 5th | % 2nd to 5th | 6th to 10th | % 6th to 10th | 11th and Lower | % 11th and Lower |
| Elias J. Corey | 1990 | 97 | 1 | 25 | 1 | 1 | 100.0% | 0 | 0.0% | 0 | 0.0% | 0 | 0.0% |
| Alan J. Heeger | 2000 | 123 | 5 | 28 | 3 | 1 | 20.0% | 4 | 80.0% | 0 | 0.0% | 0 | 0.0% |
| Osamu Shimomura | 2008 | 50 | 1 | 29 | 1 | 1 | 100.0% | 0 | 0.0% | 0 | 0.0% | 0 | 0.0% |
| Ada E. Yonath | 2009 | 36 | 3 | 27.5 | 1 | 2 | 66.7% | 1 | 33.3% | 0 | 0.0% | 0 | 0.0% |
| Ei-ichi Negishi | 2010 | 49 | 6 | 7 | 2 | 6 | 100.0% | 0 | 0.0% | 0 | 0.0% | 0 | 0.0% |
| **PERCENT TIMES LAUREATES WERE PRIMARY AUTHOR OF THE REVIEW ARTICLES = 68.8%** ||||||||||||||

being "the first and only book that provides chemical information on all known bioluminescence systems, in a single volume" (http://books.google.com/books/about/Bioluminescence.html?id=DNrTfH5PcWoC). The other book was *Handbook of Organopalladium Chemistry for Organic Synthesis* edited by Ei-ichi Negishi (New York: Wiley-Interscience, 2002.  2 volumes).  This book is described as having contributions from "over 24 world authorities in the field" and "organized to provide maximum utility to the bench synthetic chemist" (http://books.google.com/books?id=mTMA2hExAaIC&source=gbs_navlinks_s).  It certainly fits in with Negishi's productivity in review articles.  Given the role played by the Nobelists in writing review articles and producing other works summarizing the findings of their field, the evidence is that Garfield was certainly correct in his assessment of the importance of review journals and, conversely, Narin misunderstood and miscalculated the "influence" of these journals.  As a personal aside, I once presented Garfield with the Narin assessment of review journals and was curtly dismissed with the statement: "Nobelists write review articles."  The statement was an epiphany for me.

**Conclusion**

The evidence of this paper is conclusive.  Garfield's theory of citation indexing and Google's PageRank algorithm lead to similar results, because both are based upon the same premise, i.e., that subject sets of relevant documents are defined semantically better by linkages than by words.  Garfield (1955) first set forth these premises in his seminal *Science* article entitled, "Citation Indexes for Science: A New Dimension through Association of Ideas."  As a result, Garfield's theory of citation indexing and Google's PageRank algorithm validate each other, and it is possible to use Google Scholar to evaluate the significance of the publications of researchers.